%% file: paper.tex
\documentclass[sigconf]{acmart}

\usepackage{booktabs}
\usepackage{tabularx}
\usepackage{tabulary}
\usepackage{mdwlist}
\usepackage{todonotes}
\usepackage{subcaption}
\usepackage{listings}
\usepackage{multirow}
\usepackage{makecell}
\usepackage{url}
\usepackage[T1]{fontenc}
\usepackage{balance}

\presetkeys{todonotes}{inline}{}

\newcommand \redbold[1]{\textcolor{red}{\textbf{#1}}}
\newcommand \bluebold[1]{\textcolor{blue}{\textbf{#1}}}

\begin{document}

\copyrightyear{2018}
\acmYear{2018}
\setcopyright{iw3c2w3}
\acmConference[WWW 2018]{The 2018 Web Conference}{April 23--27, 2018}{Lyon, France}
\acmBooktitle{WWW 2018: The 2018 Web Conference, April 23--27, 2018, Lyon, France}
\acmPrice{}
\acmDOI{10.1145/3178876.3186090}
\acmISBN{978-1-4503-5639-8/18/04}

\fancyhead{}

\title{Large-Scale Analysis of Style Injection\\by Relative Path Overwrite}

\author{Sajjad Arshad}
\affiliation{\institution{Northeastern University}}
\email{arshad@ccs.neu.edu}

\author{Seyed Ali Mirheidari}
\affiliation{\institution{University of Trento}}
\email{seyedali.mirheidari@unitn.it}

\author{Tobias Lauinger}
\affiliation{\institution{Northeastern University}}
\email{p672@tobias.lauinger.name}

\author{Bruno Crispo}
\affiliation{\institution{University of Trento}}
\email{bruno.crispo@unitn.it}

\author{Engin Kirda}
\affiliation{\institution{Northeastern University}}
\email{ek@ccs.neu.edu}

\author{William Robertson}
\affiliation{\institution{Northeastern University}}
\email{wkr@ccs.neu.edu}

\renewcommand{\shortauthors}{S. Arshad et al.}

\input{abstract}


\maketitle

\input{introduction}

\input{background}

\input{methodology}

\input{analysis}

\input{mitigation}

\input{conclusion}

\begin{acks}

This work was supported by the National Science Foundation (NSF) under grant CNS-1703454
award, and Secure Business Austria.

\end{acks}

\bibliographystyle{ACM-Reference-Format}
\balance
\bibliography{paper}

\end{document}

%% file: abstract.tex
\begin{abstract}

Relative Path Overwrite (RPO) is a recent technique to inject style directives
into sites even when no style sink or markup injection vulnerability is
present. It exploits differences in how browsers and web servers interpret
relative paths (i.e., \textit{path confusion}) to make a HTML page reference
itself as a stylesheet; a simple text injection vulnerability along with
browsers' leniency in parsing CSS resources results in an attacker's ability to
inject style directives that will be interpreted by the browser. Even though
style injection may appear less serious a threat than script injection, it has
been shown that it enables a range of attacks, including secret exfiltration.

In this paper, we present the first large-scale study of the Web to measure the
prevalence and significance of style injection using RPO. Our work shows that
around 9\,\% of the sites in the Alexa Top 10,000 contain at
least one vulnerable page, out of which more than one third can be exploited. We
analyze in detail various impediments to successful exploitation, and make
recommendations for remediation. In contrast to script injection, relatively
simple countermeasures exist to mitigate style injection. However, there appears
to be little awareness of this attack vector as evidenced by a range of popular
Content Management Systems (CMSes) that we found to be exploitable.

\end{abstract}

%% file: introduction.tex
\section{Introduction}

Cross-Site Scripting (XSS)~\cite{owasp_xss} attacks are one of the most common
threats on the Web. While XSS has traditionally been understood as the
attacker's capability to inject script into a site and have it executed by
the victim's web browser, more recent work has shown that script injection is
not a necessary precondition for effective attacks. By injecting Cascading Style
Sheet (CSS) directives, for instance, attackers can carry out so-called
\textit{scriptless} attacks~\cite{ccs2012scriptless} and exfiltrate secrets from
a site.

The aforementioned injection attacks typically arise due to the lack of
separation between code and data~\cite{ccs2013dedacota}, and more specifically,
insufficient sanitization of untrusted inputs in web applications. While script
injection attacks are more powerful than those based on style injection, they
are also more well-known as a threat, and web developers are comparatively more
likely to take steps to make them more difficult. From an attacker's point of
view, style injection attacks may be an option in scenarios where script
injection is not possible.

There are many existing techniques of how style directives could be injected
into a site~\cite{ccs2012scriptless,ccs2010cross_origin_css}. A relatively
recent class of attacks is Relative Path Overwrite (RPO), first proposed in a
blog post by Gareth Heyes~\cite{rpo} in 2014. These attacks exploit the semantic
disconnect between web browsers and web servers in interpreting relative paths
(\textit{path confusion}). More concretely, in certain settings an attacker can
manipulate a page's URL in such a way that the web server still returns the
same content as for the benign URL. However, using the manipulated URL as the
base, the web browser incorrectly expands relative paths of included resources,
which can lead to resources being loaded despite not being intended to be
included by the developer. Depending on the implementation of the site,
different variations of RPO attacks may be feasible. For example, an attacker
could manipulate the URL to make the page include user-generated content
hosted on the same domain~\cite{rpo_techniques}. When an injection vulnerability
is present in a page, an attacker could manipulate the URL such that the web
page references itself as the stylesheet, which turns a simple text injection
vulnerability into a style sink~\cite{rpo}. Among these attack instantiations,
the latter variant has preconditions that are comparatively frequently met by
sites. Our work focuses on this variant of RPO.

To date, little is known about how widespread RPO vulnerabilities are on the
Web. Especially since the attack is more recent and less well-known than
traditional XSS, we believe it is important to characterize the extent of the
threat and quantify its enabling factors. In this paper, we present the first
in-depth study of style injection vulnerability using RPO. We extract pages
using relative-path stylesheets from the Common Crawl
dataset~\cite{common_crawl}, automatically test if style directives can be
injected using RPO, and determine whether they are interpreted by the browser.
Out of 31 million pages from 222 thousand Alexa Top 1\,M
sites~\cite{alexa_top_1m} in the Common Crawl that use relative-path
stylesheets, we find that 377\,k pages (12\,k sites) are vulnerable; 11\,k
pages on 1\,k sites can be exploited in Chrome, and nearly 55\,k pages on
over 3\,k sites can be exploited in Internet Explorer. We analyze a range of
factors that prevent a vulnerable page from being exploited, and discuss how
these could be used to mitigate these vulnerabilities.

The contributions of this paper are summarized as follows:

\begin{itemize}

\item We present the first automated and large-scale study of the prevalence and
significance of RPO vulnerabilities in the wild.

\item We discuss a range of factors that prevent a vulnerability from being
exploited, and find that simple countermeasures exist to mitigate RPO.

\item We link many exploitable pages to installations of Content Management
Systems (CMSes), and notify the vendors.

\end{itemize}

%% file: background.tex
\section{Background \& Related Work}
\label{sec:background}

The general threat model of Relative Path Overwrite (RPO) resembles that of
Cross-Site Scripting (XSS). Typically, the attacker's goal is to steal sensitive
information from a third-party site or make unauthorized transactions on the
site, such as gaining access to confidential financial information or
transferring money out of a victim's account.

The attacker carries out the attack against the site indirectly, by way of a
victim that is an authorized user of the site. The attacker can trick the victim
into following a crafted link, such as when the victim visits a domain under the
attacker's control and the page automatically opens the manipulated link, or
through search engine poisoning, deceptive shortened links, or through means of
social engineering.

\subsection{Cross-Site Scripting}

Many sites have vulnerabilities that let attackers inject malicious script.
Dynamic sites frequently accept external inputs that can be controlled by an
attacker, such as data in URLs, cookies, or forms. While the site developer's
aim would have been to render the input as text, lack of proper sanitization can
result in the input being executed as script~\cite{owasp_xss_defense}. The
inclusion of unsanitized inputs could occur on the server side or client side,
and in a persistent \textit{stored} or volatile \textit{reflected}
way~\cite{owasp_xss}. To the victim's web browser, the code appears as
originating from the first-party site, thus it is given full access to the
session data in the victim's browser. Thereby, the attacker bypasses protections
of the Same-Origin Policy.

\subsection{Scriptless Attacks}

Cross-Site Scripting is perhaps the most well-known web-based attack, against
which many sites defend by filtering user input. Client-side security mechanisms
such as browser-based XSS filters~\cite{www2010xss_auditor} and Content Security
Policy~\cite{www2010csp,w3c_csp} also make it more challenging for attackers to
exploit injection vulnerabilities for XSS. This has led attackers (and
researchers) to investigate potential alternatives, such as \textit{scriptless}
attacks. These attacks allow sniffing users' browsing
histories~\cite{dsn2014scriptless_timing,w2sp2010history_sniffing}, exfiltrating
arbitrary content~\cite{css_font_face}, reading HTML
attributes~\cite{sexy_assassin_css,csp_nonce_bypass}, and bypassing Clickjacking
defenses~\cite{sexy_assassin_css}. In the following, we highlight two types of
scriptless attacks proposed in the literature. Both assume that an attacker
cannot inject or execute script into a site. Instead, the attacker abuses
features related to Cascading Style Sheets (CSS).

Heiderich et al.~\cite{ccs2012scriptless} consider scenarios where an attacker
can inject CSS into the context of the third-party page so that the style
directives are interpreted by the victim's browser when displaying the page.
That is, the injection sink is either located inside a style context, or the
attacker can inject markup to create a style context around the malicious CSS
directives. While the CSS standard is intended for styling and layout purposes
such as defining sizes, colors, or background images and as such does not
contain any traditional scripting capabilities, it does provide some
context-sensitive features that, in combination, can be abused to extract and
exfiltrate data. If the secret to be extracted is not displayed, such as a token
in a hidden form field or link URL, the attacker can use the CSS attribute
accessor and content property to extract the secret and make it visible as text,
so that style directives can be applied to it. Custom attacker-supplied fonts
can change the size of the secret text depending on its value. Animation
features can be used to cycle through a number of fonts in order to test
different combinations. Media queries or the appearance of scrollbars can be
used to implement conditional style, and data exfiltration by loading a
different URL for each condition from the attacker's server. Taken together,
Heiderich et al. demonstrate that these techniques allow an attacker to steal
credit card numbers or CSRF tokens~\cite{owasp_csrf_defense} without script
execution.

Rather than using layout-based information leaks to exfiltrate data from a page,
Huang et al.~\cite{ccs2010cross_origin_css} show how syntactically lax parsing
of CSS can be abused to make browsers interpret an HTML page as a
``stylesheet.'' The attack assumes that the page contains two injection sinks,
one before and one after the location of the secret in the source code. The
attacker injects two CSS fragments such as
\texttt{\{\}*\{background:url('//attacker.com/?} and \texttt{');\}}, which make
the secret a part of the URL that will be loaded from the attacker's server when
the directive is interpreted. It is assumed that the attacker cannot inject
markup, thus the injected directive is not interpreted as style when the site
is conventionally opened in a browser. However, the CSS standard mandates that
browsers be very forgiving when parsing CSS, skipping over parts they do not
understand~\cite{css_syntax}. In practice, this means that an attacker can set
up a site that loads the vulnerable third-party site \textit{as a
stylesheet}. When the victim visits the attacker's site while logged in, the
victim's browser loads the third-party site and interprets the style directive,
causing the secret to be sent to the attacker. To counter this attack, modern
browsers do not load documents with non-CSS content types and syntax errors as
stylesheets when they originate from a different domain than the including page.
Yet, attacks based on tolerant CSS parsing are still feasible when both the
including and the included page are loaded from the same domain. Relative Path
Overwrite attacks can abuse such a scenario~\cite{rpo_gadgets}.

\subsection{Relative Path Overwrite}
\label{sec:background:rpo}

Relative Path Overwrite vulnerabilities can occur in sites that use relative
paths to include resources such as scripts or stylesheets. Before a web browser
can issue a request for such a resource to the server, it must expand the
relative path into an absolute URL. For example, assume that a web browser has
loaded an HTML document from \url{http://example.com/rpo/test.php} which
references a remote stylesheet with the relative path \url{dist/styles.css}. Web
browsers treat URLs as file system-like paths, that is, \url{test.php} would be
assumed to be a file within the parent directory \url{rpo/}, which would be used
as the starting point for relative paths, resulting in the absolute URL
\url{http://example.com/rpo/dist/styles.css}.

However, the browser's interpretation of the URL may be very different from how
the web server resolves the URL to determine which resource should be returned
to the browser. The URL may not correspond to an actual server-side file system
structure at all, or the web server may internally rewrite parts of the URL. For
instance, when a web server receives a request for
\url{http://example.com/rpo/test.php/} with an added trailing slash, it may
still return the same HTML document corresponding to the \url{test.php}
resource. Yet, to the browser this URL would appear to designate a directory
(without a file name component), thus the browser would request the stylesheet
from \url{http://example.com/rpo/test.php/dist/styles.css}. Depending on the
server configuration, this may either result in an error since no such file
exists, or the server may interpret \url{dist/styles.css} as a parameter to the
script \url{test.php} and return the HTML document. In the latter case, the HTML
document includes itself as a stylesheet. Provided that the document contains a
(text) injection vulnerability, attackers can carry out the scriptless attacks;
since the stylesheet inclusion is same-origin, the document load is permitted.

The first account of RPO is attributed to a blog post by Gareth
Heyes~\cite{rpo}, introducing self-referencing a PHP script with server-side URL
rewriting. Furthermore, the post notes that certain versions of Internet
Explorer allow JavaScript execution from within a CSS context in the
\textit{Compatibility View} mode~\cite{cvlist}, escalating style injection to
XSS~\cite{csscripting}. Another blog post by Dalili~\cite{rpo_iis} extends the
technique to IIS and ASP.Net applications, and shows how URL-encoded slashes are
decoded by the server but not the browser, allowing not only self-reference but
also the inclusion of different resources. Kettle~\cite{prssi} coins the term
Path Relative StyleSheet Import (PRSSI) for a specific subset of RPO attacks,
introduces a PRSSI vulnerability scanner for Burp Suite~\cite{burpsuite}, and
proposes countermeasures. Terada~\cite{rpo_techniques} provides more
exploitation techniques for various browsers or certain web applications, and
\cite{rpo_gadgets} discusses an example chaining several vulnerabilities to
result in a combination of RPO and a double style injection attack.
Gil shows how attackers can deceive web cache servers by using
RPO~\cite{blackhatusa2017webcache,webcache_deception}. Some of the
attacks discussed in the various blog posts are custom-tailored to specific
sites or applications, whereas others are more generic and apply to certain
web server configurations or frameworks.

\subsection{Preconditions for RPO Style Attacks}

For the purpose of this paper, we focus on a generic type of RPO attack because
its preconditions are less specific and are likely met by a larger number of
sites. More formally, we define a page as \textit{vulnerable} if:

\begin{itemize}

\item The page includes at least one stylesheet using a relative path.

\item The server is set up to serve the same page even if the URL is manipulated
by appending characters that browsers interpret as path separators.

\item The page reflects style directives injected into the URL or cookie. Note
that the reflection can occur in an arbitrary location within the page, and
markup or script injection are not necessary.

\item The page does not contain a \texttt{<base>} HTML tag before relative paths
that would let the browser know how to correctly expand them.

\end{itemize}

This attack corresponds to style injection by means of a page that references
itself as a stylesheet (PRSSI). Since the ``stylesheet'' self-reference is, in
fact, an HTML document, web servers would typically return it with a
\texttt{text/html} content type. Browsers in standards-compliant mode do not
attempt to parse documents with a content type other than CSS even if referenced as
a stylesheet, causing the attack to fail. However, web browsers also support
\textit{quirks mode} for backwards compatibility with non-standards compliant
sites~\cite{browser_modes_doctype}; in this mode, browsers ignore the content
type and parse the document according to the inclusion context only.

We define a vulnerable page as \textit{exploitable} if the injected style is
interpreted by the browser--that is, if an attacker can force browsers to render
the page in quirks mode. This can occur in two alternative ways:

\begin{itemize}

\item The vulnerable HTML page specifies a \textit{document type} that causes
the browser to use quirks mode instead of standards mode. The document type
indicates the HTML version and dialect used by the page;
Section~\ref{sec:analysis:doctypes} provides details on how the major web
browsers interpret the document types we encountered during our study.

\item Even if the page specifies a document type that would usually result in
standards mode being used, quirks mode parsing can often be enforced in Internet
Explorer~\cite{prssi}. Framed documents inherit the parsing mode from the parent
document, thus an attacker can create an attack page with an older document type
and load the vulnerable page into a frame. This trick only works in Internet
Explorer, however, and it may fail if the vulnerable page uses any anti-framing
technique, or if it specifies an explicit value for the \texttt{X-UA-Compatible}
HTTP header (or equivalent).

\end{itemize}

Our measurement methodology in Section~\ref{sec:methodology} tests how often
these preconditions hold in the wild in order to quantify the vulnerability and
exploitability of pages with respect to RPO attacks.

\input{related_work.tex}

%% file: related_work.tex
\subsection{Related Work}
\label{sec:related_work}

In the previous sections, we surveyed a number of style-based attacks in the
scientific literature, and several blog posts discussing special cases of
RPO. We are not aware of any scholarly work about RPO, or any research about how
prevalent RPO vulnerabilities are on the Web. To the best of our knowledge, Burp
Suite~\cite{burpsuite} is the first and only tool that can detect PRSSI
vulnerabilities based on RPO in web applications. However, in contrast to our
work, it does not determine if the vulnerability can be exploited. Furthermore,
we are the first to provide a comprehensive survey of how widespread RPO style
vulnerabilities and exploitabilities are in the wild.

The separate class of script-based attacks has been studied extensively, such as
systematic analysis of XSS sanitization frameworks~\cite{esorics2011xss},
detecting XSS vulnerabilities in Rich Internet
Applications~\cite{asiaccs2012flashover}, large-scale detection of DOM-based
XSS~\cite{ccs2013domxss,blackhatasia2014domxss}, and bypassing XSS mitigations
by Script Gadgets~\cite{blackhatusa2017script_gadgets,ccs2017script_gadgets}. An
array of XSS prevention mechanisms have been proposed, such as XSS
Filter~\cite{xss_filter}, XSS-Guard~\cite{dimva2008xss_guard},
SOMA~\cite{ccs2008soma}, BluePrint~\cite{sp2009blueprint}, Document Structure
Integrity~\cite{ndss2009doc_integrity}, XSS Auditor~\cite{www2010xss_auditor},
NoScript~\cite{noscript}, Context-Sensitive Auto-Sanitization
(CSAS)~\cite{ccs2011csas}, DOM-based XSS filtering using runtime taint
tracking~\cite{usenixsec2014client_side_xss}, preventing script injection
through software design~\cite{commacm2014tangled_web}, Strict
CSP~\cite{ccs2016cspisdead}, and DOMPurify~\cite{esorics2017dompurify}.
However, the vulnerability measurements and proposed countermeasures of these
works on script injection do not apply to RPO-based style injection.

%% file: methodology.tex
\section{Methodology}
\label{sec:methodology}

Our methodology consists of three main phases. We seed our system with pages
from the Common Crawl archive to extract \textit{candidate} pages that include
at least one stylesheet using a relative path. To determine whether these
candidate pages are \textit{vulnerable}, we attempt to inject style directives
by requesting variations of each page's URL to cause \textit{path confusion} and
test whether the generated response reflects the injected style directives.
Finally, we test how often vulnerable pages can be \textit{exploited} by
checking whether the reflected style directives are parsed and used for
rendering in a web browser.

\subsection{Candidate Identification}
\label{sec:methodology:candidate}

For finding the initial seed set of candidate pages with relative-path
stylesheets, we leverage the Common Crawl from August 2016, which contains more
than 1.6 billion pages. By using an existing dataset, we can quickly
identify candidate pages without creating any web crawl traffic. We use a Java
HTML parser to filter any pages containing only inline CSS or stylesheets
referenced by absolute URLs, leaving us with over 203 million pages on nearly 6
million sites. For scalability purposes, we further reduce the set of
candidate pages in two steps:

\begin{enumerate}

\item We retain only pages from sites listed in the Alexa Top 1 million
ranking, which reduces the number of candidate pages to 141 million pages on 223
thousand sites. In doing so, we bias our result toward popular sites--that is,
sites where attacks could have a larger impact because of the higher number of
visitors.

\item We observed that many sites use templates customized through query
strings or path parameters. We expect these templates to cause similar
vulnerability and exploitability behavior for their instantiations, thus we can
speed up our detection by grouping URLs using the same template, and testing
only one random representative of each group.

In order to group pages, we replace all the values of query parameters with
constants, and we also replace any number identifier in the path with a
constant. We group pages that have the same abstract URL as well as the same
document type in the Common Crawl dataset. For example, we would group
\url{example.com/?lang=en} and \url{example.com/?lang=fr}.

\end{enumerate}

Since our methodology contains a step during which we actively test whether a
vulnerability can be exploited, we remove from the candidate set all pages
hosted on sites in \texttt{.gov}, \texttt{.mil}, \texttt{.army}, \texttt{.navy},
and \texttt{.airforce}. The final candidate set consists of 137 million pages
(31 million page groups) on 222 thousand sites.

\subsection{Vulnerability Analysis}
\label{sec:methodology:vulnerable}

To determine whether a candidate page is vulnerable, we implemented a
lightweight crawler based on the Python Requests module. At a high level, the
crawler simulates how a browser expands relative paths and tests whether style
directives can be injected into the resources loaded as stylesheets using path
confusion.

For each page group from the candidate set, the crawler randomly selects one
representative URL and mutates it according to a number of techniques explained
below. Each of these techniques aims to cause path confusion and taints page
inputs with a style directive containing a long unique, random string. The
crawler requests the mutated URL from the server and parses the response
document, ignoring resources loaded in frames. If the response contains a
\texttt{<base>} tag, the crawler considers the page not vulnerable since the
\texttt{<base>} tag, if used correctly, can avoid path confusion. Otherwise, the
crawler extracts all relative stylesheet paths from the response and expands
them using the mutated URL of the main page as the base, emulating how browsers
treat relative paths (see Section~\ref{sec:background:rpo}). The crawler then
requests each unique stylesheet URL until one has been found to reflect the
injected style in the response.

The style directive we inject to test for reflection vulnerabilities is shown in
the legend of Figure~\ref{fig:taint-techniques}. The payload begins with an
encoded newline character, as we observed that the presence of a newline
character increases the probability of a successful injection. We initially use
\texttt{\%0A} as the newline character, but also test \texttt{\%0C} and
\texttt{\%0D} in case of unsuccessful injection. The remainder of the payload
emulates the syntax of a simple CSS directive and mainly consists of a randomly
generated string used to locate the payload in the body of the server response.
If the crawler finds a string match of the injected unique string, it considers
the page vulnerable.

In the following, we describe the various URL mutation techniques we use to
inject style directives. All techniques also use RPO so that instead of the
original stylesheet files, browsers load different resources that are more
likely to contain an injection vulnerability. Conceptually, the RPO approaches
we use assume some form of server-side URL rewriting as described in
Section~\ref{sec:background:rpo}. That is, the server internally resolves a
crafted URL to the same script as the ``clean'' URL. Under that assumption, the
path confusion caused by RPO would result in the page referencing itself as
the stylesheet when loaded in a web browser. However, this assumption is only
conceptual and not necessary for the attack to succeed. For servers that do not
internally rewrite URLs, our mutated URLs likely cause error responses since the
URLs do not correspond to actual files located on these servers. Error responses
are typically HTML documents and may contain injection sinks, such as when they
display the URL of the file that could not be found. As such, server-generated
error responses can be used for the attack in the same way as regular pages.

Our URL mutation techniques differ in how they attempt to cause path confusion
and inject style directives by covering different URL conventions used by a
range of web application platforms.

\input{figures/vulnerability_analysis}

\paragraph{\textbf{Path Parameter}}

A number of web frameworks such as PHP, ASP, or JSP can be configured to use URL
schemes that encode script input parameters as a directory-like string following
the name of the script in the URL. Figure~\ref{fig:taint:parameter-simple} shows
a generic example where there is no parameter in the URL. Since the crawler does
not know the name of valid parameters, it simply appends the payload as a
subdirectory to the end of the URL. In this case, content injection can occur if
the page reflects the page URL or referrer into the response. Note that in
the example, we appended two slashes so that the browser does not remove the
payload from the URL when expanding the stylesheet reference to the parent
directory (\url{../style.css}). In the actual crawl, we always appended twenty
slashes to avoid having to account for different numbers of parent directories.
We did not observe relative paths using large numbers of \url{../} to
reference stylesheets, thus we are confident that twenty slashes suffice for our
purposes.

Different web frameworks handle path parameters slightly differently, which is
why we distinguish a few additional cases. If parameters are present in the URL,
we can distinguish these cases based on a number of regular expressions that we
generated. For example, parameters can be separated by slashes
(Figure~\ref{fig:taint:parameter-php}, PHP or ASP) or semicolons (
Figure~\ref{fig:taint:parameter-jsp}, JSP). When the crawler detects one of
these known schemes, it injects the payload into each parameter. Consequently, in
addition to URL and referrer reflection, injection can also be successful when
any of the parameters is reflected in the page.

\paragraph{\textbf{Encoded Path}}

This technique targets web servers such as IIS that decode encoded slashes in
the URL for directory traversal, whereas web browsers do not. Specifically, we
use \texttt{\%2F}, an encoded version of `\texttt{/}', to inject our payload
into the URL in such a way that the canonicalized URL is equal to the original
page URL (see Figure~\ref{fig:taint:path}). Injection using this technique
succeeds if the page reflects the page URL or referrer into its output.

\paragraph{\textbf{Encoded Query}}

Similar to the technique above, we replace the URL query delimiter `\texttt{?}'
with its encoded version \texttt{\%3F} so that web browsers do not interpret it
as such. In addition, we inject the payload into every value of the query
string, as can be seen in Figure~\ref{fig:taint:query}. CSS injection happens if
the page reflects either the URL, referrer, or any of the query values in
the HTML response.

\paragraph{\textbf{Cookie}}

Since stylesheets referenced by a relative path are located in the same origin
as the referencing page, its cookies are sent when requesting the stylesheet. CSS
injection may be possible if an attacker can create new cookies or tamper with
existing ones (a strong assumption compared to the other techniques), and if the
page reflects cookie values in the response. As shown in
Figure~\ref{fig:taint:cookie}, the URL is only modified by adding slashes to
cause path confusion. The payload is injected into each cookie value and sent by
the crawler as an HTTP header.

\subsection{Exploitability Analysis}
\label{sec:methodology:exploitable}

Once a page has been found to be vulnerable to style injection using RPO, the
final step is to verify whether the reflected CSS in the response is evaluated
by a real browser. To do so, we built a crawler based on Google Chrome, and used
the Remote Debugging Protocol~\cite{debugging_protocol} to drive the browser and
record HTTP requests and responses. In addition, we developed a Chrome extension
to populate the cookie header in CSS stylesheet requests with our payload.

In order to detect exploitable pages, we crawled all the pages from the previous
section that had at least one reflection. Specifically, for each page we checked
which of the techniques in Figure~\ref{fig:taint-techniques} led to reflection,
and crafted the main URL with a CSS payload. The CSS payload used to verify
exploitability is different from the simple payload used to test reflection.
Specifically, the style directive is prefixed with a long sequence of
\texttt{\}} and \texttt{]} characters to close any preceding open curly braces
or brackets that may be located in the source code of the page, since they might
prevent the injected style directive from being parsed correctly. The style
directive uses a randomly-generated URL to load a background image for the HTML
body. We determine whether the injected style is evaluated by checking the
browser's network traffic for an outgoing HTTP request for the image.

\paragraph{\textbf{Overriding Document Types}}

Reflected CSS is not always interpreted by the browser. One possible explanation
is the use of a modern document type in the page, which does not cause the
browser to render the page in quirks mode. Under certain circumstances, Internet
Explorer allows a parent page to force the parsing mode of a framed page into
quirks mode~\cite{prssi}. To test how often this approach succeeds in practice,
we also crawled vulnerable pages with Internet Explorer~11 by framing them
while setting \texttt{X-UA-Compatible} to \texttt{IE=EmulateIE7} via a
\texttt{meta} tag in the attacker's page.

\subsection{Limitations}

RPO is a class of attacks and our methodology covers only a subset of them. We
target RPO for the purpose of style injection using an HTML page referencing
itself (or, accidentally, an error page) as the stylesheet. In terms of style
injection, our crawler only looks for reflection, not stored injection of style
directives. Furthermore, manual analysis of a site might reveal more
opportunities for style injection that our crawler fails to detect
automatically.

For efficiency reasons, we seed our analysis with an existing Common Crawl
dataset. We do not analyze the vulnerability of pages not contained in the
Common Crawl seed, which means that we do not cover all sites, and we do not
fully cover all pages within a site. Consequently, the results presented in
this paper should be seen as a lower bound. If desired, our methodology can be
applied to individual sites in order to analyze more pages.

\subsection{Ethical Considerations}
\label{sec:methodology:ethics}

One ethical concern is that the injected CSS might be stored on the server
instead of being reflected in the response, and it could break sites as a
result. We took several cautionary steps in order to minimize any damaging side
effects on sites we probed. First, we did not try to login to the site,
and we only tested RPO on the publicly available version of the page. In
addition, we only requested pages by tainting different parts of the URL,
and did not submit any forms. Moreover, we did not click on any button or link
in the page in order to avoid triggering JavaScript events. These steps
significantly decrease the chances that injected CSS will be stored on the
server. In order to minimize the damaging side effects in case our injected CSS
was stored, the injected CSS is not a valid style directive, and even if it is
stored on the server, it will not have any observable effect on the page.

In addition, experiment resulted in the discovery of vulnerable content
management systems (CMSes) used world-wide, and we contacted them so they can
fix the issue. We believe the real-world experiments that we conducted were
necessary in order to measure the risk posed by these vulnerabilities and inform
site owners of potential risks to their users.

%% file: figures/vulnerability_analysis.tex
\begin{figure}[t]
\centering
{\sffamily

\begin{subfigure}[t]{1\columnwidth}
\begin{lstlisting}[basicstyle=\fontsize{8}{8}\ttfamily]
/page.asp
/page.asp@@/@@**PAYLOAD**@@//@@
/page.asp@@/@@**PAYLOAD**@@/@@style.css
\end{lstlisting}
\caption{Path Parameter (Simple)}
\label{fig:taint:parameter-simple}
\end{subfigure}

\begin{subfigure}[t]{1\columnwidth}
\begin{lstlisting}[basicstyle=\fontsize{8}{8}\ttfamily]
/page.php/param1/param2
/page.php/**PAYLOAD**param1/**PAYLOAD**param2@@//@@
/page.php/**PAYLOAD**param1/**PAYLOAD**param2@@/@@style.css
\end{lstlisting}
\caption{Path Parameter (PHP or ASP)}
\label{fig:taint:parameter-php}
\end{subfigure}

\begin{subfigure}[t]{1\columnwidth}
\begin{lstlisting}[basicstyle=\fontsize{8}{8}\ttfamily]
/page.jsp;param1;param2
/page.jsp;**PAYLOAD**param1;**PAYLOAD**param2@@//@@
/page.jsp;**PAYLOAD**param1;**PAYLOAD**param2@@/@@style.css
\end{lstlisting}
\caption{Path Parameter (JSP)}
\label{fig:taint:parameter-jsp}
\end{subfigure}

\begin{subfigure}[t]{1\columnwidth}
\begin{lstlisting}[basicstyle=\fontsize{8}{8}\ttfamily]
/dir/page.aspx
/**PAYLOAD**@@/..%2F@@dir/**PAYLOAD**@@/..%2F@@page.aspx@@//@@
/**PAYLOAD**@@/..%2F@@dir/**PAYLOAD**@@/..%2F@@page.aspx@@/@@style.css
\end{lstlisting}
\caption{Encoded Path}
\label{fig:taint:path}
\end{subfigure}

\begin{subfigure}[t]{1\columnwidth}
\begin{lstlisting}[basicstyle=\fontsize{8}{8}\ttfamily]
/page.html?k1=v1&k2=v2
/page.html@@%3F@@k1=**PAYLOAD**v1&k2=**PAYLOAD**v2@@//@@
/page.html@@%3F@@k1=**PAYLOAD**v1&k2=**PAYLOAD**v2@@/@@style.css
\end{lstlisting}
\caption{Encoded Query}
\label{fig:taint:query}
\end{subfigure}

\begin{subfigure}[t]{1\columnwidth}
\begin{lstlisting}[basicstyle=\fontsize{8}{8}\ttfamily]
/page.php?key=value
/page.php@@//@@?key=value
/page.php@@/@@style.css

Original Cookie: k1=v1; k2=v2
 Crafted Cookie: k1=**PAYLOAD**v1; k2=**PAYLOAD**v2
\end{lstlisting}
\caption{Cookie}
\label{fig:taint:cookie}
\end{subfigure}

\caption{Various techniques of \redbold{path confusion} and \bluebold{style
injection}. In each example, the first URL corresponds to the regular page, and
the second one to the page URL crafted by the attacker. Each HTML page is
assumed to reference a stylesheet at \url{../style.css}, resulting in the
browser expanding the stylesheet path as shown in the third URL.
\bluebold{PAYLOAD} corresponds to
\bluebold{\texttt{\%0A\{\}body\{background:NONCE\}}} (simplified), where
\texttt{NONCE} is a randomly generated string.}

\label{fig:taint-techniques}
}
\end{figure}

%% file: analysis.tex
\section{Analysis}
\label{sec:analysis}

For the purposes of our analysis, we gradually narrow down the seed data from
the Common Crawl to pages using relative style paths in the Alexa Top 1\,M,
reflecting injected style directives under RPO, and being exploitable due to
quirks mode rendering.

Table~\ref{tab:dataset_stats} shows a summary of our dataset.
\textit{Tested Pages} refers to the set of randomly selected pages
from the page groups as discussed in Section~\ref{sec:methodology:candidate}.
For brevity, we are referring to \textit{Tested Pages} wherever we mention
pages in the remainder of the paper.

\subsection{Relative Stylesheet Paths}
\label{sec:analysis:relative}

\input{tables/dataset.tex}

\begin{figure*}
\centering
\begin{minipage}{.32\textwidth}
    \includegraphics[width=1\textwidth,height=.8\textwidth]{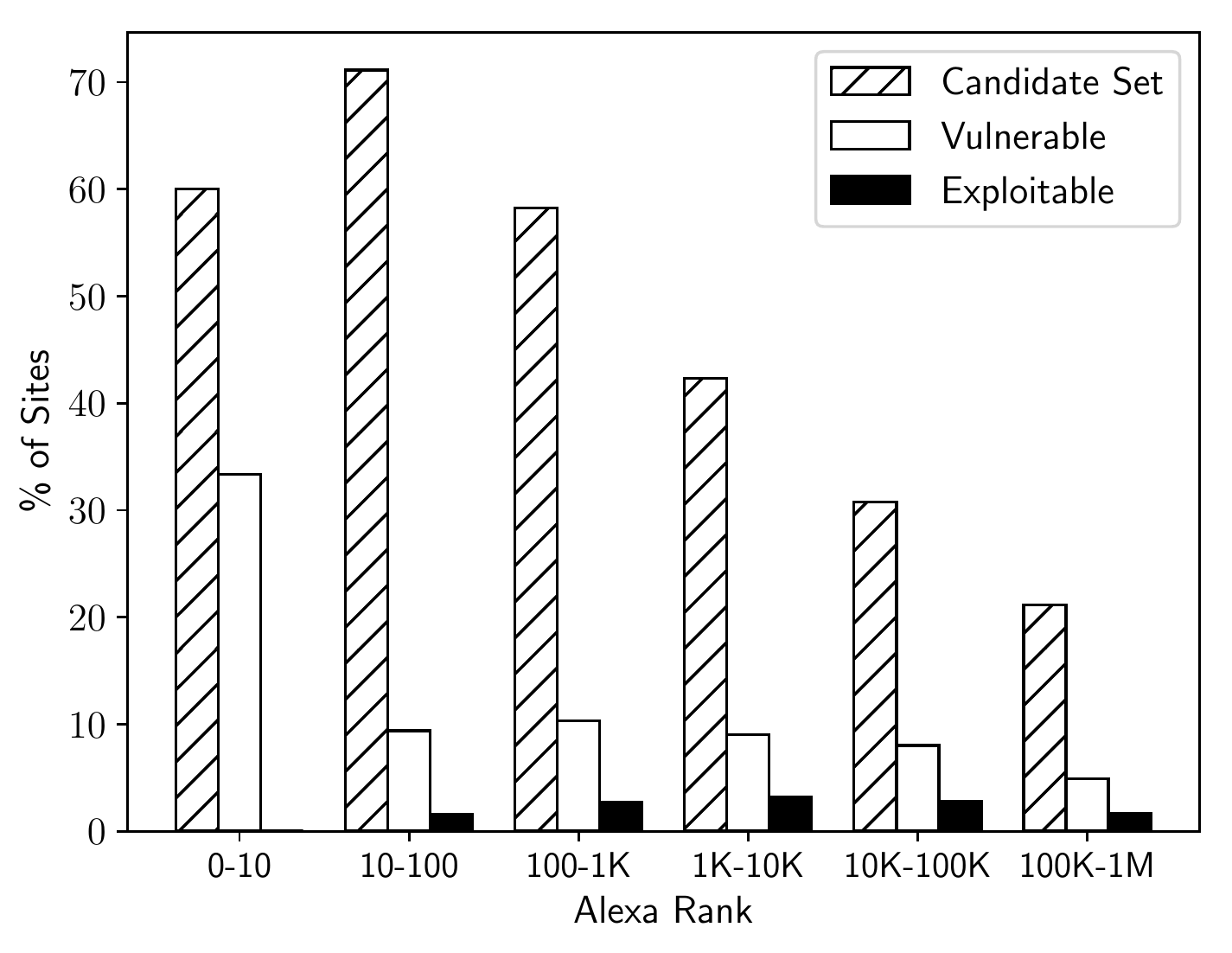}
    \caption{Percentage of the Alexa site ranking in our candidate set
             (exponentially increasing bucket size).}
    \label{fig:analysis:alexa_rank}
\end{minipage}
\hfill
\begin{minipage}{.32\textwidth}
    \includegraphics[width=1\textwidth,height=.8\textwidth]{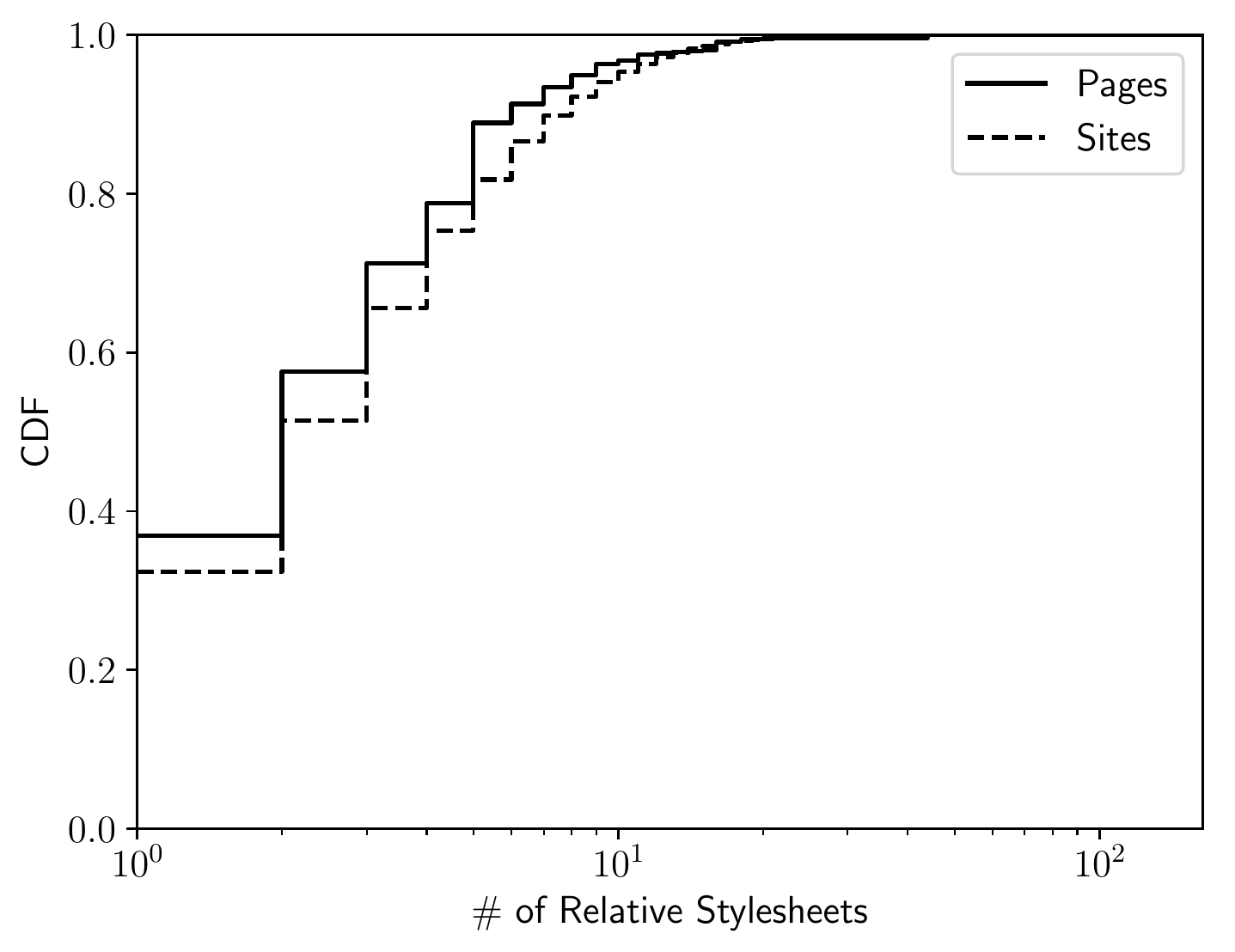}
    \caption{CDF of total and maximum number of relative stylesheets per web
             page and site, respectively.}
    \label{fig:analysis:relative_stylesheets}
\end{minipage}
\hfill
\begin{minipage}{.32\textwidth}
    \includegraphics[width=1\textwidth,height=.8\textwidth]{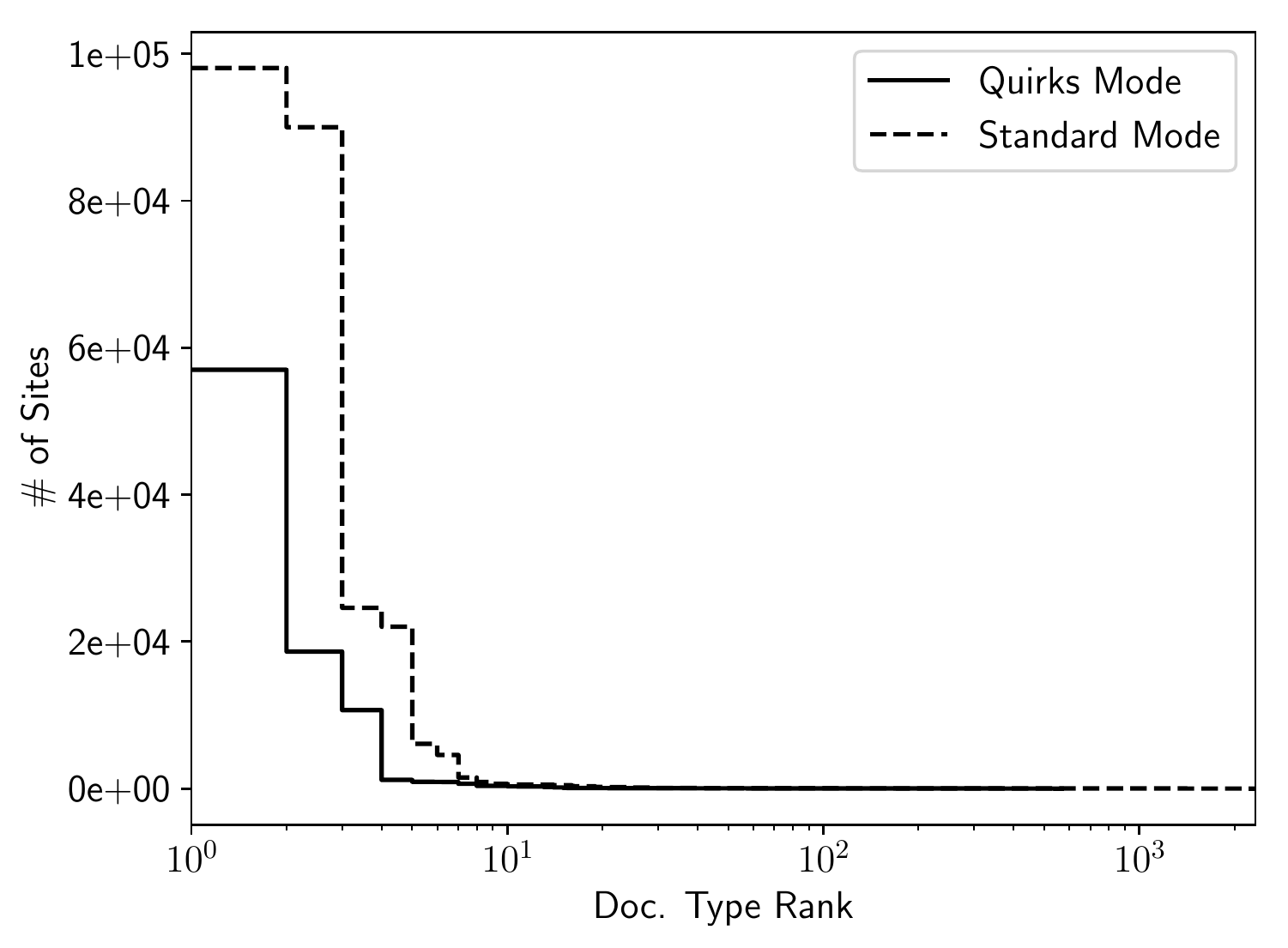}
    \caption{Number of sites containing at least one page with a certain
             document type (ordered by doctype rank).}
    \label{fig:analysis:doctypes_rank_sites}
\end{minipage}
\end{figure*}

To assess the extent to which our Common Crawl-seeded candidate set covers sites
of different popularity, consider the hatched bars in
Figure~\ref{fig:analysis:alexa_rank}. Six out of the ten largest sites according
to Alexa are represented in our candidate set. That is, they are contained in
the Common Crawl, and have relative style paths. The figure shows that our
candidate set contains a higher fraction of the largest sites and a lower
fraction of the smaller sites. Consequently, our results better represent the
most popular sites, which receive most visitors, and most potential victims of
RPO attacks.

While all the pages in the candidate set contain at least one relative
stylesheet path, Figure \ref{fig:analysis:relative_stylesheets} shows that
63.1\,\% of them contain multiple relative paths, which increases the chances of
finding a successful RPO and style injection point.

\subsection{Vulnerable Pages}
\label{sec:analysis:vulnerable}

\input{tables/vulnerable_exploitable_results}

We consider a candidate page vulnerable if one of the style injection techniques
of Section~\ref{sec:methodology:vulnerable} succeeds. In other words, the
server's response should reflect the injected payload. Furthermore, we
conservatively require that the response not contain a \texttt{base} tag since a
correctly configured base tag can prevent path confusion.

Table~\ref{tab:vulnerable_exploitable_result} shows that 1.2\,\% of pages
are vulnerable to at least one of the injection techniques, and 5.4\,\% of
sites contain at least one vulnerable page. The path parameter technique is
most effective against pages, followed by the encoded query and the encoded path
techniques. Sites that are ranked higher according to Alexa are more likely
to be vulnerable, as shown in Figure~\ref{fig:analysis:alexa_rank}, where
vulnerable and exploitable sites are relative to the candidate set in each
bucket. While one third of the candidate set in the Top~10 (two out of six
sites) is vulnerable, the percentage oscillates between 8 and 10\,\% among the
Top~100\,k. The candidate set is dominated by the smaller sites in the ranks
between 100\,k and 1\,M, which have a vulnerability rate of 4.9\,\% and push
down the average over the entire ranking.

A \texttt{base} tag in the server response can prevent path confusion because it
indicates how the browser should expand relative paths. We observed a number of
inconsistencies with respect to its use. At first, 603 pages on 60 sites
contained a \texttt{base} tag in their response; however, the server response
after injecting our payload did not contain the tag anymore, rendering these
pages potentially exploitable. Furthermore, Internet Explorer's implementation
of the \texttt{base} tag appears to be broken. When such a tag is present,
Internet Explorer fetches two URLs for stylesheets---one expanded according to
the base URL specified in the tag, and one expanded in the regular, potentially
``confused'' way of using the page URL as the base. In our experiments, Internet
Explorer always applied the ``confused'' stylesheet, even when the one based on
the \texttt{base} tag URL loaded faster. Consequently, \texttt{base} tags do not
appear to be an effective defense against RPO in Internet Explorer (They seem to
work as expected in other browsers, including Edge).

\subsection{Exploitable Pages}
\label{sec:analysis:exploitable}

To test whether a vulnerable page was exploitable, we opened it in Chrome,
injected a style payload with an image reference (randomly generated URL), and
checked if the image was indeed loaded. This test succeeded for 2.9\,\% of
vulnerable pages; 0.5\,\% of sites in the candidate set had at least one
exploitable page (Table~\ref{tab:vulnerable_exploitable_result}).

In the following, we explore various factors that may impact whether a
vulnerable page can be exploited, and we show how some of these partial defenses
can be bypassed.

\subsubsection{Document Types}
\label{sec:analysis:doctypes}

\input{tables/browsers_doctypes.tex}

\input{tables/top_quirksmode_doctypes.tex}

HTML document types play a significant role in RPO-based style injection attacks
because browsers typically parse resources with a non-CSS content type in a CSS
context only when the page specifies an ancient or non-standard HTML document
type (or none at all). The pages in our candidate set contain a total of 4,318
distinct document types. However, the majority of these unique document types
are not standardized and differ from the standardized ones only by small
variations, such as forgotten spaces or misspellings.

To determine how browsers interpret these document types (i.e., whether they
cause them to render a page in standards or quirks mode), we designed a
controlled experiment. For each unique document type, we set up a local page
with a relative stylesheet path and carried out an RPO attack to inject CSS
using a payload similar to what we described in
Section~\ref{sec:methodology:vulnerable}. We automatically opened the local page
in Chrome, Firefox, Edge, Internet Explorer, Safari, and Opera, and we kept
track of which document type caused the injected CSS to be parsed and the
injected background image to be downloaded.

Table~\ref{tab:doctypes_browsers} contains the results of this experiment. Even
though the exact numbers vary among browsers, roughly a third of the unique
document types we encountered result in quirks mode rendering. Not surprisingly,
both Microsoft products Edge and Internet Explorer exhibit identical results,
whereas the common Webkit ancestry of Chrome, Opera, and Safari also show
identical results. Overall, 1,271 (29.4\,\%) of the unique document types force
all the browsers into quirks mode, whereas 1,378 (31.9\,\%) of them cause at
least one browser to use quirks mode rendering.
Table~\ref{tab:top_quirksmode_doctypes} shows the most frequently used document
types that force all the browsers into quirks mode, which includes the absence
of a document type declaration in the page.

To test how often Internet Explorer allows a page's document type to be
overridden when loading it in an \texttt{iframe}, we created another controlled
experiment using a local attack page framing the victim page, as outlined in
Section~\ref{sec:methodology:exploitable}. Using Internet Explorer~11, we loaded
our local attack page for each unique document type inside the frame, and tested
if the injected CSS was parsed. While Internet Explorer parsed the injected CSS
for 1,319 (30.5\,\%) of the document types in the default setting, the frame
override trick caused CSS parsing for 4,248 (98.4\,\%) of the unique document
types.

While over one thousand document types result in quirks mode, and around three
thousand document types cause standards mode parsing, the number of document
types that have been standardized is several orders of magnitude smaller. In
fact, only a few (standardized) document types are used frequently in pages,
whereas the majority of unique document types are used very rarely.
Figure~\ref{fig:analysis:doctypes_rank_sites} shows that only about ten
standards and quirks mode document types are widely used in pages and
sites. Furthermore, only about 9.6\,\% of pages in the candidate set use
a quirks mode document type; on the remaining pages, potential RPO style
injection vulnerabilities cannot be exploited because the CSS would not be
parsed (unless Internet Explorer is used). However, when grouping pages in the
candidate set by site, 32.2\,\% of sites contain at least one page
rendered in quirks mode (Table~\ref{tab:doctypes_summary}), which is one of the
preconditions for successful RPO.

\input{tables/doctypes_summary.tex}

\subsubsection{Internet Explorer Framing}

We showed above that by loading a page in a frame, Internet Explorer can be
forced to disregard a standards mode document type that would prevent
interpretation of injected style. To find out how often this technique can be
applied for successful RPO attacks, we replicated our Chrome experiment in
Internet Explorer, this time loading each vulnerable page inside a frame. Around
14.5\,\% of vulnerable pages were exploitable in Internet Explorer, five times
more than in Chrome (1.6\,\% of the sites in the candidate set).

Figure~\ref{fig:analysis:alexa_rank} shows the combined exploitability results
for Chrome and Internet Explorer according to the rank of the site. While our
methodology did not find any exploitable vulnerability on the six highest-ranked
sites in the candidate set, between 1.6\,\% and 3.2\,\% of candidate sites in each
remaining bucket were found to be exploitable. The highest exploitability rate
occurred in the ranks 1\,k through 10\,k.

Broken down by injection technique, the framing trick in Internet Explorer
results in more exploitable pages for each technique except for cookie injection
(Table~\ref{tab:vulnerable_exploitable_result}). One possible explanation for
this difference is that the Internet Explorer crawl was conducted one month
after the Chrome crawl, and sites may have changed in the meantime.
Furthermore, we observed two additional impediments to successful exploitation
in Internet Explorer that do not apply to Chrome. The framing technique is
susceptible to frame-busting methods employed by the framed pages, and Internet
Explorer implements an anti-MIME-sniffing header that Chrome appears to ignore.
We analyze these issues below.

\subsubsection{Anti-Framing Techniques}

Some sites use a range of techniques to prevent other pages from loading
them in a frame~\cite{w2sp2010frame_busting}. One of these techniques is the
\texttt{X-Frame-Options} header. It accepts three different values:
\texttt{DENY}, \texttt{SAMEORIGIN}, and \texttt{ALLOW-FROM} followed by a
whitelist of URLs.

In the vulnerable dataset, 4,999 pages across 391 sites use this header
correctly and as a result prevent the attack. However, 1,900 pages across 34
sites provide incorrect values for this header, and we successfully attack
552 pages on 2 sites with Internet Explorer.

A related technique is the \texttt{frame-ancestors} directive provided by
Content Security Policy. It defines a (potentially empty) whitelist of URLs
allowed to load the current page in a frame, similar to \texttt{ALLOW-FROM}.
However, it is not supported by Internet Explorer, thus it cannot be used to
prevent the attack.

Furthermore, developers may use JavaScript code to prevent framing of a page.
Yet, techniques exist to bypass this
protection~\cite{owasp_clickjacking_defence}. In addition, the attacker can use
the HTML 5 \texttt{sandbox} attribute in the \texttt{iframe} tag and omit the
\texttt{allow-top-navigation} directive to render JavaScript frame-busting code
ineffective. However, we did not implement any of these techniques to allow
framing, which means that more vulnerable pages could likely be exploited in
practice.

\subsubsection{MIME Sniffing}

A consequence of self-reference in the type of RPO studied in this paper is that
the HTTP content type of the fake ``stylesheet'' is \texttt{text/html} rather
than the expected \texttt{text/css}. Because many sites contain misconfigured
content types, many browsers attempt to infer the type based on the request
context or file extension (\textit{MIME sniffing}), especially in quirks mode.
In order to ask the browser to disable content sniffing and refuse interpreting
data with an unexpected or wrong type, sites can set the header
\texttt{X-Content-Type-Options: nosniff}
~\cite{sp2009contentsniff,firefox_mime_sniff,content_type_options}.

To determine whether the injected CSS is still being parsed and executed in
presence of this header while the browser renders in quirks mode, we ran an
experiment similar to Section~\ref{sec:analysis:doctypes}. For each browser in
Table~\ref{tab:doctypes_browsers}, we extracted the document types in which the
browser renders in quirks mode, and for each of them, we set up a local page
with a relative stylesheet path. We then opened the page in the browser,
launched an RPO attack, and monitored if the injected CSS was executed.

Only Firefox, Internet Explorer, and Edge respected this header and did not
interpret injected CSS in any of the quirks mode document types. The
remaining browsers did not block the stylesheet even though the content type was
not \texttt{text/css}. With an additional experiment, we confirmed that Internet
Explorer blocked our injected CSS payload when \texttt{nosniff} was set, even in
the case of the framing technique.

Out of all the vulnerable pages, 96,618 pages across 232 sites had a
\texttt{nosniff} response header; 23 pages across 10 sites were confirmed
exploitable in Chrome but not in Internet Explorer, since the latter browser
respects the header while the former does not.

\subsection{Content Management Systems}
\label{sec:analysis:cmses}

While analyzing the exploitable pages in our dataset, we noticed that many
appeared to belong to well-known CMSes. Since these web applications are
typically installed on thousands of sites, fixing RPO weaknesses in these
applications could have a large impact.

To identify CMSes, we visited all exploitable pages using
Wappalyzer~\cite{wappalyzer}. Additionally, we detected two CMSes that were not
supported by Wappalyzer. Overall, we identified 23 CMSes on 41,288 pages
across 1,589 sites. Afterwards, we manually investigated whether the RPO
weakness stemmed from the CMS by installing the latest version of each CMS (or
using the online demo), and testing whether exploitable paths found in our
dataset were also exploitable in the CMS. After careful analysis, we confirmed
four CMSes to be exploitable in their most recent version that are being used by
40,255 pages across 1,197 sites.

Out of the four exploitable CMSes, one declares no document type and one uses a
quirks mode document type. These two CMSes can be exploited in Chrome, whereas
the remaining two can be exploited with the framing trick in Internet Explorer.
Beyond the view of our Common Crawl candidate set, Wappalyzer detected nearly
32\,k installations of these CMSes across the Internet, which suggests that many
more sites could be attacked with RPO. We reported the RPO weaknesses to the
vendors of these CMSes using recommended notification techniques
~\cite{usenixsec2016vulnnotify1,usenixsec2016vulnnotify2,weis2017vulnnotify}.
Thus far, we heard back from one of the vendors, who acknowledged the
vulnerability and are going to take the necessary steps to fix the issue.
However, we have not received any response from the other vendors.

%% file: tables/dataset.tex
\begin{table}[t]
    \caption{Narrowing down the Common Crawl to the candidate set used in our analysis
    (from left to right).}
    \label{tab:dataset_stats}
    \centering
	\footnotesize
    \begin{tabular}{lrrr}
    \toprule
    \multicolumn{2}{r}{\textbf{Relative CSS}} & \textbf{Alexa Top 1M} & \textbf{Candidate Set} \\
    \midrule
    All Pages & 203,609,675 & 141,384,967 & 136,793,450 \\
    Tested Pages & 53,725,270 & 31,448,446 & 30,991,702 \\
    Sites & 5,960,505 & 223,212 & 222,443 \\
    Doc. Types & 9,833 & 2,965 & 2,898 \\
    \bottomrule
    \end{tabular}
\end{table}

%% file: tables/vulnerable_exploitable_results.tex
\begin{table*}[t]
    \caption{Vulnerable/exploitable pages and sites in the
             candidate set (IE using framing).}
    \label{tab:vulnerable_exploitable_result}
    \centering
    \footnotesize
    \begin{tabular}{lrrrrrr}
    \toprule
    \multirow{2}{*}{\textbf{Technique}} &
    \multicolumn{2}{c}{\textbf{Vulnerable}} &
    \multicolumn{2}{c}{\textbf{Exploitable (Chrome)}} &
    \multicolumn{2}{c}{\textbf{Exploitable (Internet Explorer)}} \\

    \cmidrule[0.5pt](lr){2-3}
    \cmidrule[0.5pt](lr){4-5}
    \cmidrule[0.5pt](lr){6-7}

    &
    \textbf{Pages} &
    \textbf{Sites} &
    \textbf{Pages} &
    \textbf{Sites} &
    \textbf{Pages} &
    \textbf{Sites}
    \\

    \midrule

    Path Parameter & 309,079 (1.0\%) & 9,136 (4.1\%) & 6,048 (<0.1\%) & 1,025 (0.5\%) & 52,344 (0.2\%) & 3,433 (1.5\%) \\
    Encoded Path & 53,502 (0.2\%) & 1,802 (0.8\%) & 3 (<0.1\%) & 2 (<0.1\%) & 24 (<0.1\%) & 5 (<0.1\%) \\
    Encoded Query & 89,757 (0.3\%) & 1,303 (0.6\%) & 23 (<0.1\%) & 20 (<0.1\%) & 137 (<0.1\%) & 43 (<0.1\%) \\
    Cookie & 15,656 (<0.1\%) & 1,030 (0.5\%) & 4,722 (<0.1\%) & 81 (<0.1\%) & 2,447 (<0.1\%) & 238 (0.1\%) \\

    \midrule

    Total & 377,043 (1.2\%) & 11,986 (5.4\%) & 10,781 (<0.1\%) & 1,106 (0.5\%) & 54,853 (0.2\%) & 3,645 (1.6\%) \\

    \bottomrule
    \end{tabular}

\end{table*}

%% file: tables/browsers_doctypes.tex
\begin{table}[t]
    \caption{Quirks mode document types by browser.}
    \label{tab:doctypes_browsers}
    \centering
    \footnotesize
    \begin{tabular}{lllr}
    \toprule

    \textbf{Browser} &
    \textbf{Version} &
    \textbf{Operating System} &
    \textbf{Doc. Types} \\

    \midrule
    Chrome & 55 & Ubuntu 16.04 & 1,378 (31.9\,\%) \\
    Opera & 42 & Ubuntu 16.04 & 1,378 (31.9\,\%) \\
    Safari & 10 & macOS Sierra & 1,378 (31.9\,\%) \\
	\addlinespace
    Firefox & 50 & Ubuntu 16.04 & 1,326 (30.7\,\%) \\
	\addlinespace
    Edge & 38 & Windows 10 & 1,319 (30.5\,\%) \\
    Internet Explorer & 11 & Windows 7 & 1,319 (30.5\,\%) \\
    \bottomrule
    \end{tabular}
\end{table}

%% file: tables/top_quirksmode_doctypes.tex
\begin{table}[t]
    \caption{Most frequent document types causing all browsers to render
    in quirks mode, as well as the sites that use at least one such
    document type.}
    \label{tab:top_quirksmode_doctypes}
    \centering
	\footnotesize
    \setlength{\tabcolsep}{4pt}
    \begin{tabular}{lrr}
    \toprule
    \textbf{Doc. Type (shortened)} & \textbf{Pages} & \textbf{Sites} \\
    \midrule
    (none) & 1,818,595 (5.9\,\%) & 56,985 (25.6\,\%) \\
    "-//W3C//DTD HTML 4.01 Transitional//EN" & 721,884 (2.3\,\%) & 18,648 (8.4\,\%) \\
    "-//W3C//DTD HTML 4.0 Transitional//EN" & 385,656 (1.2\,\%) & 11,566 (5.2\,\%) \\
    "-//W3C//DTD HTML 3.2 Final//EN" & 22,019 (<0.1\,\%) & 1,175 (0.5\,\%) \\
    "-//W3C//DTD HTML 3.2//EN" & 10,839 (<0.1\,\%) & 927 (0.4\,\%) \\
    \midrule
    All & 3,046,449 (9.6\,\%) & 71,597 (32.2\,\%) \\
    \bottomrule
    \end{tabular}
\end{table}

%% file: tables/doctypes_summary.tex
\begin{table}[t]
    \caption{Summary of document type usage in sites.}
    \label{tab:doctypes_summary}
    \centering
    \footnotesize
    \begin{tabular}{lrr}
    \toprule
    \textbf{Doc. Type} & \textbf{At Least One Crawled Page} & \textbf{All Crawled Pages} \\
    \midrule

    None & 56,985 (25.6\%) & 19,968 (9.0\%) \\
    Quirks & 27,794 (12.5\%) & 7,720 (3.5\%) \\
    None or Quirks & 71,597 (32.2\%) & 30,040 (13.5\%) \\
    \addlinespace
    Standards & 192,403 (86.5\%) & 150,846 (67.8\%) \\
    
    \bottomrule
    \end{tabular}
\end{table}

%% file: mitigation.tex
\section{Mitigation Techniques}
\label{sec:mitigation}

Relative path overwrites rely on the web server and the web browser interpreting
URLs differently. HTML pages can use only absolute (or root-relative) URLs,
which removes the need for the web browser to expand relative paths.
Alternatively, when the HTML page contains a \texttt{<base>} tag, browsers are
expected to use the URL provided therein to expand relative paths instead of
interpreting the current document's URL. Both methods can remove ambiguities and
render RPO impossible if applied correctly. Specifically, base URLs must be set
according to the server's content routing logic. If developers choose to
calculate base URLs dynamically on the server side rather than setting them
manually to constant values, there is a risk that routing-agnostic algorithms
could be confused by manipulated URLs and re-introduce attack opportunities by
instructing browsers to use an attacker-controlled base URL. Furthermore,
Internet Explorer does not appear to implement this tag correctly.

Web developers can reduce the attack surface of their sites by eliminating
any injection sinks for strings that could be interpreted as a style directive.
However, doing so is challenging because in the attack presented in this paper,
style injection does not require a specific sink type and does not need the
ability of injecting markup. Injection can be accomplished with relatively
commonly used characters, that is, alphanumeric characters and
\texttt{()\{\}/"}. Experience has shown that despite years of efforts, even
context-sensitive and more special character-intensive XSS injection is still
possible in many sites, which leads us to believe that style injection will
be similarly difficult to eradicate. Even when all special characters in user
input are replaced by their corresponding HTML entities and direct style
injection is not possible, more targeted RPO attack variants referencing
existing files may still be feasible. For instance, it has been shown that
user uploads of seemingly benign profile pictures can be used as ``scripts'' (or
stylesheets)~\cite{rpo_techniques}.

Instead of preventing RPO and style injection vulnerabilities, the most
promising approach could be to avoid exploitation. In fact, declaring a modern
document type that causes the HTML document to be rendered in standards mode
makes the attack fail in all browsers except for Internet Explorer. Web
developers can harden their pages against the frame-override technique in
Internet Explorer by using commonly recommended HTTP headers:
\texttt{X-Content-Type-Options} to disable ``content type sniffing'' and always
use the MIME type sent by the server (which must be configured correctly),
\texttt{X-Frame-Options} to disallow loading the page in a frame, and
\texttt{X-UA-Compatible} to turn off Internet Explorer's compatibility view.

%% file: conclusion.tex
\section{Conclusion}
\label{sec:conslusion}

This paper presented a systematic study of CSS injection by RPO in the wild. We
showed that over 5\,\% of sites in the intersection of the Common Crawl and
the Alexa Top 1M are vulnerable to at least one injection technique. While the
number of exploitable sites depends on the browser and is much smaller in
relative terms, it is still consequential in absolute terms with thousands of
affected sites. RPO is a class of attacks, and our automated crawler tested
for only a subset of conceivable attacks. Therefore, the results of our study
should be seen as a lower bound; the true number of exploitable sites is
likely higher.

Compared to XSS, it is much more challenging to avoid injection of style
directives. Yet, developers have at their disposal a range of simple mitigation
techniques that can prevent their sites from being exploited in modern
browsers.